\documentclass[a4paper]{llncs}

\usepackage{alltt}
\usepackage{caption}
\usepackage{paralist}
\usepackage{nicefrac}
\usepackage{booktabs}
\usepackage{algorithm}
\usepackage{algorithmic}
\usepackage{xspace}
\usepackage{amsmath}
\usepackage{marvosym}
\usepackage{wasysym}
\usepackage{verbatim}
\usepackage{subfigure}
\usepackage{hyperref}
\usepackage{multirow}
\usepackage{cite}
\usepackage[capitalize]{cleveref}
\usepackage[per-mode=symbol,detect-all]{siunitx}
\usepackage{graphics}
\usepackage{amssymb}
\usepackage{wrapfig}
\usepackage{enumitem}

\usepackage{tikz,ifthen,pgfplots}
\usetikzlibrary{arrows,trees,backgrounds,automata,shapes,decorations,plotmarks,fit,calc,positioning,shadows,chains}
\tikzstyle{every pin edge}=[<-,shorten <=1pt]
\tikzstyle{neuron}=[circle,fill=black!25,minimum size=17pt,inner sep=0pt]
\tikzstyle{input neuron}=[neuron, fill=green!50]
\tikzstyle{output neuron}=[neuron, fill=red!50]
\tikzstyle{hidden neuron}=[neuron, fill=blue!50]
\tikzstyle{annot} = [text width=4em, text centered]
\tikzstyle{nnedge} = [-{stealth},shorten >=0.1cm, shorten <=0.05cm,line width=0.8pt,black]

\usetikzlibrary{calc}

\newcommand{\sat}{\texttt{SAT}}
\newcommand{\unsat}{\texttt{UNSAT}}

\begin{document}

\title{Simplifying Neural Networks using Formal Verification}

\author{
  Sumathi Gokulanathan\inst{1} \and
  Alexander Feldsher\inst{1} \and
  Adi Malca\inst{1} \and
  Clark Barrett\inst{2} \and
  Guy Katz \inst{1}\textsuperscript{(\Letter)}
}
\institute{
  The Hebrew University of Jerusalem, Jerusalem, Israel \\
  \{sumathi.giokolanat, feld, adimalca, guykatz\}@cs.huji.ac.il
  \and
  Stanford University, Stanford, USA \\
  barrett@cs.stanford.edu 
}

\maketitle

\begin{abstract}
  Deep neural network (DNN) verification is an emerging field, with
  diverse verification engines quickly becoming
  available. Demonstrating the effectiveness of these engines on
  real-world DNNs is an important step towards their wider
  adoption. We present a tool that can leverage existing verification
  engines in performing a novel application: neural network
  simplification, through the reduction of the size of a DNN without
  harming its accuracy. We report on the work-flow of the
  simplification process, and demonstrate its potential significance
  and applicability on a family of real-world DNNs for aircraft
  collision avoidance, whose sizes we were able to reduce by as much
  as 10\%.
\end{abstract}


\section{Introduction}

\emph{Deep neural networks} (\emph{DNNs}) are revolutionizing the way
complex software is produced, obtaining unprecedented results in
domains such as image recognition~\cite{SiZi14}, natural language
processing~\cite{CoWeBoKaKaKu11}, and game
playing~\cite{SiHuMaGuSiVaScAnPaLaDi16}. There is now
even a trend of using DNNs as controllers in autonomous cars and
unmanned aircraft~\cite{BoDeDwFiFlGoJaMoMuZhZhZhZi16,JuLoBrOwKo16}.
With DNNs becoming prevalent, it is highly important to develop
automatic techniques to assist in creating, maintaining and adjusting
them.

As DNNs are used in tackling increasingly complex tasks, their sizes
(i.e., number of neurons) are also increasing --- to a point where
modern DNNs can have millions of neurons~\cite{HoZhChKaWaWeAnAd17}. DNN size is thus
becoming a liability, as deploying larger networks takes up more
space, increases energy consumption, and prolongs response
times. Network size can even become a limiting factor in situations
where system resources are scarce.  For example, consider the ACAS Xu
airborne collision avoidance system for unmanned aircraft, which is
currently being developed by the Federal Aviation Administration~\cite{JuLoBrOwKo16}.
This is a highly safety-critical system, for which a DNN-based
implementation is being considered~\cite{JuLoBrOwKo16}. Because this system will be
mounted on actual drones with limited memory, efforts are being made
to reduce the sizes of the ACAS Xu DNNs as much as possible, without
harming their accuracy~\cite{JuKoOw19,JuLoBrOwKo16}.

Most work to date on DNN simplification uses various heuristics, and
does not provide formal guarantees about the simplified network's
resemblance to the original. A common approach is to start with a
large network, and reduce its size by removing some of its components
(i.e., neurons and edges)~\cite{HaMaDa15,IaHaMoAsDaKe16}. The parts to
be removed from the network are determined heuristically, and network
accuracy may be harmed, sometimes requiring additional training after
the simplification process has been performed~\cite{HaMaDa15}.


Here, we propose a novel simplification technique that harnesses
recent advances in DNN verification
(e.g.,~\cite{KaBaDiJuKo17Reluplex,GeMiDrTsChVe18,WaPeWhYaJa18}). Using
verification queries, we propose to identify components of the network
that \emph{never affect its output}. These components can be safely removed,
creating a smaller network that is completely equivalent to the
original.  We empirically demonstrate that many such removable
components exist in networks of interest.

We implement our technique in a proof-of-concept tool, called
\emph{NNSimplify}. The tool uses the following work-flow:
\begin{inparaenum}[(i)]
\item it performs lightweight simulations to identify parts of the
  DNN that are candidates for removal;
\item it invokes an underlying verification engine to dispatch
  queries that determine which of those parts can
  indeed be removed without affecting the network's outputs; and
\item it constructs the simplified network, which is equivalent to the
  original.
\end{inparaenum}
A major benefit of the proposed verification-based simplification is
that it does not require any retraining of the simplified network,
which may be expensive.

Our implementation of NNSimplify (available
online~\cite{NNSimplifyCode}) can use existing DNN verification tools
as a backend. For the evaluation reported here, we used the
recently published Marabou framework~\cite{KaHuIbJuLaLiShThWuZeDiKoBa19Marabou} as the underlying
verification engine.  We evaluated our approach on the ACAS Xu family
of DNNs for airborne collision avoidance~\cite{JuLoBrOwKo16}, and were able to reduce
the sizes of these DNNs by up to 10\% --- a highly significant reduction
for systems where resources are scarce.

The rest of the paper is organized as follows. In
Section~\ref{sec:background}, we provide a brief
background on DNNs and their verification and simplification.
Next, we describe our verification-based approach to simplification in
Section~\ref{sec:simplificationAndVerification}, followed by an
evaluation in Section~\ref{sec:evaluation}.
We then conclude in Section~\ref{sec:conclusion}.

\section{Background: DNNs, Verification and Simplification}
\label{sec:background}

\begin{figure}[htp]
  \begin{center}
    \scalebox{1} {
      \def\layersep{2.5cm}
    \begin{tikzpicture}[shorten >=1pt,->,draw=black!50, node distance=\layersep,font=\footnotesize]

      \node[input neuron] (I-1) at (0,-1) {$v_{1}$};

      \path[yshift=0.5cm] node[hidden neuron] (H-1)
      at (\layersep,-1 cm) {$v_{2}$};
      \path[yshift=0.5cm] node[hidden neuron] (H-2)
      at (\layersep,-2 cm) {$v_{3}$};

      \node[output neuron] at (2*\layersep, -1) (O-1) {$v_{4}$};

      \draw[nnedge] (I-1) -- node[above] {$1.3$} (H-1);
      \draw[nnedge] (I-1) -- node[below] {$-2$} (H-2);
      \draw[nnedge] (H-1) -- node[above] {$0.7$} (O-1);
      \draw[nnedge] (H-2) -- node[below] {$4.2$} (O-1);

      \node[annot,above of=H-1, node distance=1cm] (hl) {Hidden layer};
      \node[annot,left of=hl] {Input layer};
      \node[annot,right of=hl] {Output layer};
    \end{tikzpicture}
    }
    \captionof{figure}{A small neural network with 2 hidden nodes in
      one hidden layer. Weights are denoted over the edges. Hidden
      node values are typically determined by computing a weighted sum
      according to the weights, and then applying a non-linear
      activation function to the result.}
    \label{fig:smallDnn}
  \end{center}
\end{figure}
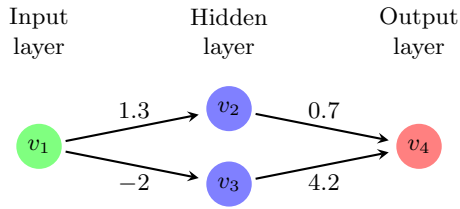

DNNs are comprised of an input layer, an output layer, and multiple
hidden layers in between. A layer is comprised of multiple nodes
(neurons), each connected to nodes from the preceding layer using a
predetermined set of weights (see Fig.~\ref{fig:smallDnn}). By assigning values to
inputs and then feeding them forward through the network, values for
each layer can be computed from the values of the previous layer,
finally resulting in values for the outputs.

As DNNs are increasingly used in safety-critical applications
(e.g.,~\cite{BoDeDwFiFlGoJaMoMuZhZhZhZi16,JuLoBrOwKo16}), there is a
surge of interest in verification methods that can provide formal
guarantees about DNN behavior. A DNN verification query consists of a
neural network and a property to be checked; and it results in either
a formal guarantee that the network satisfies the property, or a
concrete input for which the property is violated (a
counter-example). Verification queries can encode various properties
about DNNs; e.g., that slight perturbations to a network's inputs do
not affect its output, and that it is thus robust to \emph{adversarial
  perturbations}~\cite{SzZaSuBrErGoFe13,CaKaBaDi17,BaIoLaVyNoCr16}.

Recently, there has been significant progress on DNN verification
tools that can dispatch such queries (see a recent
survey~\cite{LiArLaBaKo19}).  Some of the proposed approaches for DNN
verification include the use of specialized SMT
solvers~\cite{HuKwWaWu17,KaBaDiJuKo17Reluplex,KaHuIbJuLaLiShThWuZeDiKoBa19Marabou},
the use of LP and MILP solvers~\cite{Ehlers2017,TjXiTe19}, 
symbolic interval propagation~\cite{WaPeWhYaJa18}, abstract
interpretation~\cite{GeMiDrTsChVe18}, and many others
(e.g.,~\cite{BuTuToKoKu17,DuJhSaTi18,LoMa17,NaKaRySaWa17,ElGoKa19}). This
new technology has been applied in a variety of contexts, such as
collision avoidance~\cite{KaBaDiJuKo17Reluplex}, adversarial
robustness~\cite{HuKwWaWu17,GoKaPaBa18,KaBaDiJuKo17}, hybrid
systems~\cite{SuKhSh19}, and computer networks~\cite{KaBaKaSc19}.
Although DNN verification technology is improving rapidly, scalability
remains a major limitation of existing approaches.  It has been shown
that a common variant of the DNN verification problem is NP-complete,
and becomes exponentially harder as the network size
increases~\cite{KaBaDiJuKo17Reluplex,KuKaGoJuBaKo18}.

In recent years, enormous DNNs have been appearing in order to tackle increasingly
complex tasks --- to a point where DNN size is becoming a liability, because large networks take longer to train and even to evaluate when deployed. Techniques for neural
network minimization and simplification have thus started to emerge: typically, these
take an initial, large network, and reduce its size by removing some of its components~\cite{HaMaDa15}.
The pruning phase involves the removal of edges from the network. The selection
of which edges to remove is done heuristically, often by selecting edges that have very
small weights, because these edges are less likely to significantly affect the network's outputs. If all edges connecting a node to the preceding layer or to the succeeding
layer are removed, then the node itself can be removed. After the pruning phase, the
reduced network is retrained~\cite{HaMaDa15,IaHaMoAsDaKe16}.

\section{Simplification using Verification}
\label{sec:simplificationAndVerification}

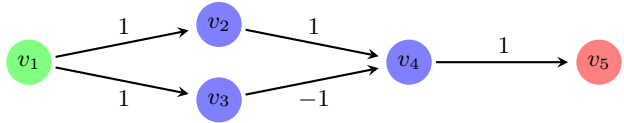
\begin{figure}[htp]
  \begin{center}
    \scalebox{1} {
      \def\layersep{2.5cm}
    \begin{tikzpicture}[shorten >=1pt,->,draw=black!50, node distance=\layersep,font=\footnotesize]

      \node[input neuron] (I-1) at (0,-1) {$v_{1}$};

      \path[yshift=0.5cm] node[hidden neuron] (H-1)
      at (\layersep,-1 cm) {$v_{2}$};
      \path[yshift=0.5cm] node[hidden neuron] (H-2)
      at (\layersep,-2 cm) {$v_{3}$};

      \node[hidden neuron] (H-3) at (2*\layersep, -1) {$v_4$};
      
      \node[output neuron] at (3*\layersep, -1) (O-1) {$v_{5}$};

      \draw[nnedge] (I-1) -- node[above] {$1$} (H-1);
      \draw[nnedge] (I-1) -- node[below] {$1$} (H-2);
      \draw[nnedge] (H-1) -- node[above] {$1$} (H-3);
      \draw[nnedge] (H-2) -- node[below] {$-1$} (H-3);
      \draw[nnedge] (H-3) -- node[above] {$1$} (O-1);

    \end{tikzpicture}
    }
    \captionof{figure}{Using verification, we can discover that node $v_4$ can safely be
      removed from the network.}
    \label{fig:simplificationExample}
  \end{center}
\end{figure}

Despite the demonstrated usefulness of pruning-based DNN
simplification~\cite{HaMaDa15,IaHaMoAsDaKe16}, heuristic-based
approaches might miss removable edges, if these edges do not have
particularly small weights. However, such edges can be identified
using verification. For example, consider the network shown in
Fig.~\ref{fig:simplificationExample}. As all edge weights have
identical magnitudes, none of them would be pruned by a
heuristic-based approach. However, using a verification engine, it is
possible to check the property: ``does there exist an input for which
$v_4$ takes a non-zero value?''. If the verification tool answers
``no'', as is the case for the network in Fig. 2 (because
$v_4 = v_2 - v_3$ and $v_2=v_3$), then we are guaranteed that $v_4$ is
always assigned 0, regardless of the input. In turn, this means that
$v_4$ can never affect nodes in subsequent layers. In this case, $v_4$
and all its edges can be safely removed from the network (rendering
the network's output constant). Due to the soundness of the
verification process, we are guaranteed that the simplified DNN is
completely equivalent to the original DNN, and thus no retraining is
required.

Using verification to identify nodes that are always
assigned 0 for every possible input, and can thus be removed, is the
core of our technique. However, because verification is costly, posing
this query for every node of the DNN might take a long time. To
mitigate this difficulty, we propose the following work-flow:

\begin{enumerate}
\item
  
  Use lightweight simulations to identify nodes that are candidates
  for removal. Initially, all hidden nodes are such candidates. We
  then evaluate the network for random input values, and remove from
  the list of candidates any hidden node that is assigned a non-zero
  value for some input. With each simulation, the number of candidates
  for removal decreases.

\item
  For each remaining candidate node $v$, we create a separate
  verification query stating that $v\neq 0$, and use the underlying
  verification engine to dispatch it. If we get an \unsat{} answer, we
  mark node $v$ for removal. The candidates are explored in a
  layer-by-layer order, which allows us to only examine a part of the
  DNN for every query. For example, when addressing a candidate in
  layer \#2, we do not encode layers \#3 and on as part of our
  verification query, as a node's assignment can only be affected
  by nodes in preceding layers. Because verifying smaller networks is
  generally easier, this layer-by-layer approach accelerates the
  process as a whole.  In addition, this process naturally lends
  itself to parallelization, by running each verification query on a
  separate machine.

\item
  Finally, we construct the simplified network, in which the nodes
  marked for removal and all their incoming and outgoing edges are
  deleted. We can also remove any nodes that subsequently become
  irrelevant due to the removal of all of their incoming or outgoing
  edges (e.g., for the DNN in Fig. 2, after removing $v_4$ we can also
  remove $v_2$ and $v_3$, as neither has any remaining outgoing
  edges).
\end{enumerate}

We note that our technique can be extended to simplify DNNs in
additional ways, by using different verification queries. For example,
it can identify separate nodes that are always assigned identical,
non-zero values (duplicates) and unify them, thus reducing the overall
number of nodes. It can also identify and remove nodes that can be
expressed as linear combinations of other nodes.

\section{Evaluation}
\label{sec:evaluation}

Our proof-of-concept implementation of the approach, called
NNSimplify, is comprised of three Python modules, one for performing
each of the aforementioned steps.  The tool is general, in two ways: (1)
it can be applied to simplify any DNN, regardless of its application
domain; and (2) it can use any DNN verification engine as a backend,
 benefiting from any future improvement in verification
technology. For our experiments we used
the Marabou~\cite{KaHuIbJuLaLiShThWuZeDiKoBa19Marabou} 
verification engine. In practice, it is required that the DNN in
question be supported by the backend verification engine --- for
example, some engines may not support certain network
topologies. Additionally, the DNN needs to be provided in a format
supported by NNSimplify; currently, the tool supports the NNet
format~\cite{NNetFormat}, and we plan to extend it to additional
formats. The tool, additional documentation, and all the benchmarks
reported in this section are available online~\cite{NNSimplifyCode}.

We evaluated NNSimplify on the ACAS Xu family of DNNs for airborne
collision avoidance~\cite{JuLoBrOwKo16}. This set contains 45 DNNs,
each with 5 input neurons, 5 output neurons, and 300 hidden neurons
spread across 6 hidden layers. The ACAS Xu networks are fully connected, and
use the ReLU activation function in each of their hidden nodes --- and
are thus supported by Marabou.

\begin{figure}[h]
  \begin{center}
      \scalebox{0.8}{
        \includegraphics[width=0.7\textwidth]{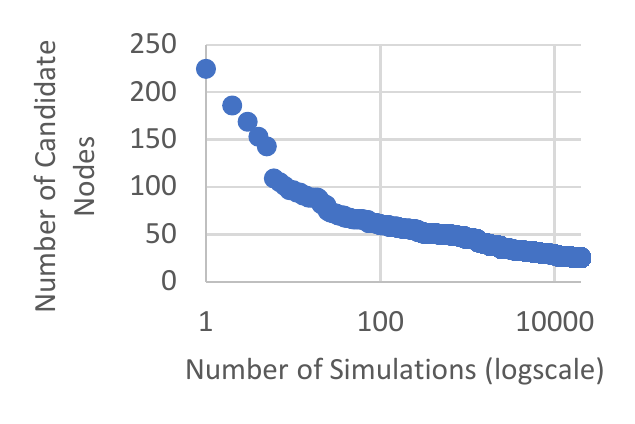}
      }
  \end{center}
  \caption{Using simulation to identify nodes that are candidates for
    removal, on one ACAS Xu network.}
  \label{fig:simulations}
\end{figure}

For each of the 45 ACAS Xu DNNs, we ran the first Python module of
NNSimplify (random simulations), resulting in a list of candidate
nodes for removal. For each DNN we performed 20000
simulations, and this narrowed down the list of nodes that are
candidates for removal to about 7\% of all hidden nodes (see
Fig.~\ref{fig:simulations}). The simulations were performed on points
sampled uniformly at random, although other distributions could of
course be used.

\begin{figure}[h]
  \begin{center}
      \scalebox{0.75}{
        \includegraphics[width=0.7\textwidth]{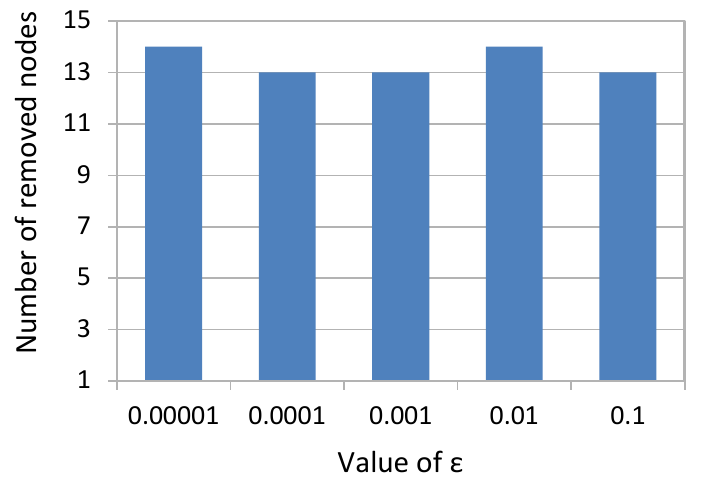}
      }
  \end{center}
  \caption{Number of removed nodes as a function of the value of
    $\varepsilon$, on one of the ACAS Xu networks.}
  \label{fig:epsilons}
\end{figure}

Next, for each candidate for removal we ran the second Python module,
which takes as input a DNN and a node $v$ that is a candidate for
removal. This module constructs a temporary, smaller DNN, where the
candidate node $v$ is the only output node (subsequent layers are
omitted). These temporary DNNs were then passed to the underlying
verification engine, with the query $v\neq 0$.  Here, we encountered
the following issue: the Marabou framework, like many
linear-programming based tools, does not provide a way to directly
specify that $v\neq 0$, but rather only to state that
$v\geq \varepsilon$ for some $\varepsilon > 0$ (we assume all hidden
nodes are, by definition, never negative, which is the case for the
ACAS Xu DNNs). We experimented with various values of $\varepsilon$
(see Fig.~\ref{fig:epsilons}), and concluded that the choice of
$\varepsilon$ has very little effect on the outcome of the experiment
--- i.e., nodes tend to either be obsolete, or take on large values.
The set of removed nodes was almost identical in all experiments, with
minor differences due to different queries timing out for different
values of $\varepsilon$.

Finally, we ran the third Python module that uses the results of the
previous steps to construct the simplified network.

\begin{figure}[h]
  \begin{center}
      \scalebox{0.9}{
        \includegraphics[width=0.7\textwidth]{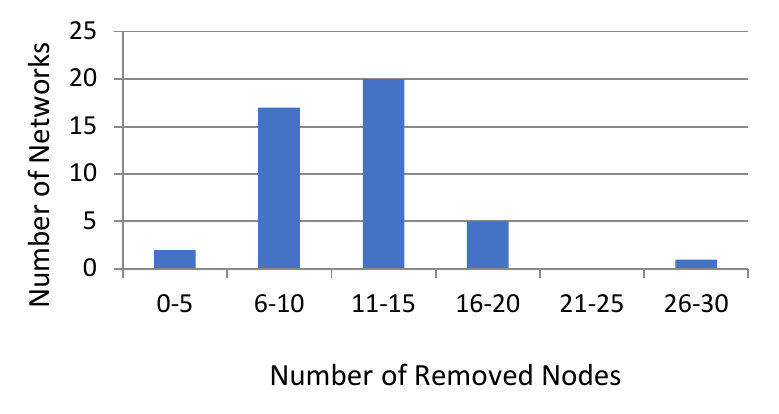}
      }
  \end{center}
  \caption{Total number of removed nodes in the ACAS Xu networks.}
  \label{fig:acasResults}
\end{figure}

We performed this process for each of the 45 DNNs. We ran
the experiments on machines with Intel Xeon E5-2670 CPUs (2.60GHz)
and 8GB of memory, and used $\varepsilon = 0.01$. Each verification
query was given a 4-hour timeout. Out of 1069 verification queries (1
per candidate node), 535 were \unsat{} (node marked for removal), 15 were
\sat{}, and 519 timed out (node not marked for removal).  Thus, on
average, 4\% of the nodes were marked for removal (535 nodes out of
13500). Fig.~\ref{fig:acasResults} depicts their distribution across the 45 DNNs. In most
networks, between 11 and 15 nodes (out of 300) could be removed; but
for a few networks, this number was higher. For one of the networks we
discovered 29 neurons that could be removed --- approximately 10\% of
that network's total number of neurons.

\section{Conclusion}
\label{sec:conclusion}

DNN verification is an emerging field, and we are just now beginning
to tap its potential in assisting engineers in DNN development. We
presented here the NNSimplify tool, which uses black-box verification
engines to simplify neural networks. We demonstrated that this
approach can lead to a substantial reduction in DNN size. Although our
experiments show that the tool is already applicable to real-world
DNNs, its scalability is limited by the scalability of its underlying
verification engine; but as the scalability of verification technology
improves, that limitation will diminish.  In the future, we plan to
extend this work along several axes. First, we intend to explore
additional verification queries, which would allow to simplify DNNs in
more sophisticated ways --- for example by revealing that some neurons
can be expressed as linear combinations of other neurons, or that some
neurons are always assigned identical values and can be merged. In
addition, we plan to investigate more aggressive simplification
steps, which may change the DNN's output, while using verification to
ensure that these changes remain within acceptable bounds. Finally,
we intend to apply the technique to additional real-world DNNs and
case studies.

\subsubsection*{Acknowledgements.}

This project was partially supported by grants from the Binational
Science Foundation (2017662), the Israel Science Foundation
(683/18), and the National Science Foundation (1814369).

{

}

\end{document}